\documentclass[twocolumn,showpacs,preprintnumbers,amsmath,amssymb]{revtex4}
\usepackage{graphicx}% Include figure files
\usepackage{dcolumn}% Align table columns on decimal point
\usepackage{bm}% bold math
\begin{document}
%\preprint{draft}
%\title{Manuscript Title:\\with Forced Linebreak}% Force line breaks with \\
%\author{Ann  Author}
 %\altaffiliation[Also at ]
 %{Physics Department,
  %XYZ University.}
%\affiliation{%
%\affiliation{
%\date{} It is always %\today, today,
%\textsc{\today{February 12}}
%\begin{abstract}
%An article usually includes an abstract, a concise summary of the work
%covered at length in the main body of the article. It is used for
%secondary publications and for information retrieval purposes. Valid
%PACS numbers may be entered using the \verb+\pacs{#1}+ command.
%\end{abstract}
%  %\keywords{Suggested keywords}%Use showkeys class option if keyword
%\maketitle
%\section{\label{sec:level1}First-level heading:\protect\\ The line
%break was forced \lowercase{via} \textbackslash\textbackslash}
\title{Principles of statistical mechanics: the energy duality}
\author{V.E. Shapiro} 
%\ead{vshapiro@triumf.ca}
\email{vshapiro@triumf.ca} 
%\today{}%February 12}
\affiliation{TRIUMF, 
%4004 Wesbrook Mall, 
Vancouver, BC, Canada}
%\date{\today}

\begin{abstract}
We argue that statistical mechanics of systems with 
relaxation implies breaking the  energy function of 
systems into two having different transformation 
rules. With this duality the energy approach incorporates 
the generalized vortex forces. 
We show general theorems and their implications and apply 
the approach to the particle confinement in fields of 
rotational symmetry. Misconceptions of extensive use of 
the quasienergy and generalized thermodynamic 
potential theories are exposed. 
\pacs{02.70.Rr \;  05.70.Ln \; 05.90.+m\;}
%\keywords{Nonequilibrium thermodynamics,  Generalized Potential,
%Vortex force, Confinement}
% \sep; stability \;relaxation  \; fluctuations}
%display desired%%\verb+\pacs{#5}+ command.
\end{abstract}
\maketitle
%\end{frontmatter}
\subsection*{Introduction}
While the main word in physics is interaction of systems 
given by an energy function,  
this notion is well defined for the ideal of closed systems 
via their motion integrals. 
For the real world, of open systems in long lasting 
non-equilibrium states
against the background of fluctuating environment, 
the meaning of energy function is vague and 
causes misleading associations and mistakes in energy reasoning and modeling. 
A new  formulation  
of energy approach to open systems is suggested in this Letter. 
 
We found this matter important even 
in conditions of low-rate irreversible forcing,   
for its  
nonlinear cumulative effect on the trend 
of system's steady state, 
stability and fluctuations can be drastic and different 
from the trend given by the  theory  of 
generalized thermodynamic potential  [1-3]
 commonly accepted in the study of 
heating/cooling, transport and phase transitions. 
This paradigm fits the ideal of closed systems, 
showing no energy vagueness.         
The generalized potential of a system relaxing in 
steady conditions to a probability density 
$\rho_{\mathrm{st}}$ connects to $\rho_{\mathrm{st}}$ 
by
\begin{equation}
\rho_{\mathrm{st}}=Ne^{-\Phi}, \quad N^{-1}=
\int e^{-\Phi}d\Gamma
\end{equation}  %      (1)
where 
the integral is over the volume  $\Gamma$ of system phase space 
and the reversible motion is on surfaces $\Phi=const$. 
The properties of the system mainly depend then on the local
properties of the minima of $\Phi$. The  analogous approach 
to systems under high frequency fields is in terms of the picture
where the hf field looks fixed or its effect is time-averaged.
In all this, Eq.~(1) merely redefines  the distribution
$\rho_{\mathrm{st}}$ in terms of function $\Phi$, and taking this
function   as the energy integral of
%\hyphenation{re-ver-sible}
reversible motion provides the physical basis of the theory, 
but implies rigid constraints.

Let us start the analysis of these constraints 
and the energy approach beyond them with formulating 
general theorems. 

\subsection*
{Three theorems: 
of entrainment, energy indeterminism and energy duality}
\texttt{I}. 
The system whose reversible-motion integral and steady 
distribution 
of states $\rho_\mathrm{st}$ are in a one-to-one relation  
must carry  along, on the average over $\rho_\mathrm{st}$, the
environment contributing to its fluctuations and dissipation.
The relation of  $\rho_\mathrm{st}$ to $\Phi$ can be taken as 
the one in 
question without loss of generality. 
Connecting $\Phi$
to the energy function of system implies
scaling this function  in terms of environmental-noise level 
of energy.  
The energy scale set so must vary proportionally with the energy 
in arbitrarily moving frames while $\Phi$ defined by (1) must 
not vary. This constraint can hold only for the systems carried 
along 
with the environment on the average and must break beyond this 
entrainment ideal, hence the proof.

\texttt{II}. 
The energy integral of reversible motion of the systems 
relaxing to a steady
distribution $\rho_\mathrm{st}$ beyond the entrainment ideal 
ceases to exist. Indeed, as follows from the proof of theorem I,  
neither 
$\Phi$ (1) nor any univalent function of it can be the energy 
integral, and it cannot be a function of transient process to  
$\rho_\mathrm{st}$ as well, for the energy integral of reversible 
motion  must be independent  of  transient states in $\Gamma$. 
 
The  energy integral break-up can be proved also as follows. 
The non-entrained steady state implies a steady mean motion 
in $\Gamma$. The relaxation to it implies that  the  
irreversible forces exerted on the motion do not vanish
as $t\rightarrow \infty$ and are of vortex type in $\Gamma$ as 
their forcing  toward the steady motion and against it differ 
in sign. 
Such forces form both the $\rho_{\mathrm{st}}$ 
and the approach to it. Thus the energy integral 
ceases to exist there, being blurred  by the  vortex forces.

\texttt{III}.
The whole blurring must come down to  
that the notion of system energy function 
(Hamiltonian) adopted as one-valued 
for a certain standing must then bifurcate.  
Its two Hamiltonians 
represent the constraints relating the partial derivatives 
of Hamiltonian with respect to the phase space variables of 
the system to its instant state. 
One is bound with $\Phi$  and
represents the constraints on the motion relative
to the environment as the source of 
diffusion/dissipation; 
the time reversal decomposition is with respect to 
the parity associated with the partial derivatives 
of just this Hamiltonian, so the irreversible drift is 
determined by them 
and hence this Hamiltonian. The other 
governs the system motion  unrelated to the said source, but 
 is also involved in forming 
the  $\rho_\mathrm{st}$ and the transition to it. 
This is the only way to comply with the arguments in I 
and II, 
and this is what we called the energy duality. 

The difference
between the two energy functions is  due to the work of
vortex forces. In its turn the difference function
acts as their energy measure. Thereby the vortex forces
are incorporated into the energy concept.

The  theorems of entrainment, indeterminism and duality
shown above admit extension to the systems in unsteady
conditions so long as  the probability description is
adequate and the probability distribution of system states
exists and relaxes from the domain of initial conditions
to a common, limit $t$-dependent distribution. The proof
follows from the fact that a  limit distribution is 
reducible to
a steady one with univalent transformations of $\Gamma$. 

Historically (e.g. reviews [3,4])  the necessary and
sufficient conditions of generalized potential theory
were formulated as detailed balance within the framework of
autonomous Fokker-Planck equations for the variables and
parameters dividable into odd and even with respect  to  
time reversal.  The detailed balance  complies with the 
entrainment ideal. Our theorems presented above set the 
energy approach to wider conditions.

Let us now give the introduced notions substance and
consider implications. 

\subsection*{Problem statement via kinetic equations}
We consider the  probability density $\rho(z,t)$ 
of system states $z$ as prime and determine 
the concept of system energy function from its time 
evolution. 
The  interactions with fluctuating environment will be
considered within the equations of general form
\begin{equation}
\frac{\partial \rho}{\partial t} =
[H,\rho] -\frac{\partial S_i}{\partial z_i}
\end{equation}  %(2)
from a given initial distribution $\rho(z,0)$ under
natural boundary conditions. Without the part 
$\partial S_i/\partial z_i$, Eq.~(2) is the 
Liouville's equation 
of a dynamic system of Hamiltonian $H$, a function of  
 a number of canonical coordinates 
$x=(x_1,\ldots,x_n)$ and moments 
$p=(p_1,\ldots,p_n)$ taken as $z$, $z=(x,p)$.
$[\,,\,]$ denotes the
Poisson brackets,
\[ [H,\rho]= \frac{\partial H}{\partial x_i}
\frac{\partial \rho }{ \partial p_i}-
\frac{\partial H}{\partial p_i} \frac{\partial \rho}
{\partial x_i}. \]
Summation over repeated dummy indexes is implied. 
$S=\{S_i\}$ is a $2n$-vector 
functional of $\rho$ in $z$, vanishing at the 
boundaries  to preserve the normalization of 
$\rho$, includes the irreversible probability 
currents. These currents depend on the 
system's motion, 
and this dependence in its turn affects the energy 
concept. 

In diffusional approximation
\begin{equation}S_i=
\left(f_i- d_{ik}\frac{\partial}
{\partial z_k}\right )\rho,\quad i,k=1,\ldots 2n
\end{equation}	%   (3)
where $f=\{f_i\}$ are dissipative forces and $d=\{d_{ik}\}$ 
is a
symmetric positive semi-definite matrix  of diffusion.
Eq.~(2) is then a general Fokker-Planck equation written in 
terms of
canonical variables of the reversible dynamics given by $H$.

$S$ may be nonlinear in $\rho$ and 
integro-differential in $z$, unless specified otherwise, and 
we use the term kinetic also for such equations (2). 
The parameters entering $H$ and $S$ may depend not only
on $z$ but also on $t$ to allow for interactions with
varying regular fields and varying chaotic environments. 

\paragraph*{Entrainment ideal.}
Its simplest modeling corresponds to  $S$ of form (3) with
$f$ and $d$ having non-vanishing components pertaining 
only to the space of $p$, with $d$ a constant 
positive-definite matrix $n\times n$ and $f$ a 
$n$-component force of  viscous friction
\begin{equation}
f_i=-\beta d_{ik} v_k
\end{equation}  %(4)
where $\beta^{-1}$  is a  scale of noise energy and
$v=(v_1,\ldots,v_n)$ the velocity given by Hamiltonian 
dynamics,
\begin{equation}
v_i=\partial H(x,p,t)/\partial p_i.
\end{equation}  % (5)
For a $t$-independent $H$ bounded below, the stationary
solution of (2) is then the Boltzmann distribution
\begin{equation}
\rho_{_{\mathrm{B}}}(z)= N e^{-\beta H(z)}, 
\quad N^{-1}= \int e^{-\beta H (z)} d^{2n}z .
\end{equation}   %(5)
The meaning of $\beta$  expounds the known equipartition 
theorem ensuing from (6):  for every component of $z$ 
(coordinate or momentum) whose contribution to $H$ 
reduces to 
a square term, say, $ \kappa_1(z_j-\kappa_2)^2$ 
with $k_1\neq 0$ 
and $\kappa_{1,2}$ independent of $z_j$, its mean 
over the Boltzmann distribution comes to
\begin{equation}
\langle \kappa_1 (z_j-\kappa_2)^2\rangle _{_{\mathrm{B}}}=
\beta^{-1}
\end{equation}   % (7)
irrespective of parameters $\kappa_{1,2}$. This  theorem
holds not only for constant $\kappa_{1,2}$ but also for
$\kappa_{1,2}$ depending on other components of $z$ 
and on
$t$, which is not insignificant for ``quasistationary''
Boltzmann distributions. By virtue of (5) and assuming
natural boundary conditions
\begin{equation}
\int v e^{-\beta H(x,p,t)}\mathrm{d}^np =0.
\end{equation}  %   (8)
So the average 
$\langle v(x,p,t)\rangle_{_{\mathrm{B}}}=0$ 
over $p$ for any $x$, 
which is the entrainment ideal for the  case.

The  same features hold  beyond the linear  friction and
constant diffusion rates, being also in effect for the  
$f$ and $d$ of elements pertaining to both $x$ and $p$
spaces and depending on  both $z$ and $t$, provided that
\begin{equation}
f_i=-\beta d_{ik}\frac{\partial H}{\partial z_k},
\quad i,k=1,\ldots2n
\end{equation}  % (9)
and that the matrix $d$ ensures approach
of the solutions of (2) to a unique steady state.  One
comes then to (6) and thus to the proof. Note that 
the $2n$ vector $\bar{v}=\partial H/ \partial z$ has 
norm $|\bar{v}|=|[z,H]|$ but $\bar{v}\bot[z,H]$.

The entrained steady state in terms of Eq.~(2) implies
\begin{equation}
(\partial S_i/\partial z_i)_{\rho=\rho_\mathrm{st}}=0.
\end{equation}  %   (10)
%for every $i=1,\ldots 2n$. 
The detailed balance defined as 
vanishing irreversible probability currents between 
any two states of the system 
complies with condition (10). Upon (10), the  
\hyphenation{equi-par-ti-tion} 
equipartition theorem loses its force beyond 
approximation (3), e.g., when jump random influences 
is a factor,  for the steady solution to (2) becomes 
non-Boltzmann, but 
$[H,\rho_\mathrm{st}(z)]$ always equals zero, 
hence, the yardstick (1). 

\subsection*{The theorem IV:  The canonical invariance
of  the irreversible operator of kinetic equations} 
Let us show that the irreversible operator of kinetic 
equations  is an invariant of canonical transformations. 
This theorem is the essence of the claimed energy 
dualism and embodies its basic rule. 

Consider a canonical transformation $z\rightarrow Z$ of 
Eq.~(2). The Poisson bracket is then to be invariant, 
as well as the
probability density $\rho(z,t)$. The latter turns into
$\Pi(Z,t)$ according to
$\rho(z,t)\mathrm{d}^{2n}z=\Pi(Z,t){d}^{2n}Z$ which
results in $\rho(z,t)=\Pi((Z(z,t),t)$   since   
the Jacobian of any canonical transformation
\begin{equation}
|\det \{\partial Z_k(z,t)/\partial z_i \}_{i,k=1}^{2n}|=1.
\end{equation}  % (11)
Under the transformation, the 
$\partial\rho/\partial t$ of (2) gives rise to
$\partial \Pi(Z,t)/\partial t$ plus the addition
$Z'_i\partial \Pi(Z,t)/\partial Z_i$ where 
\[Z'=\partial Z(z,t)/\partial t \]
is the local velocity  of map $z\rightarrow Z$.  
By virtue of (11) this velocity 
as a function of $Z$ is   
divergence-free, $\partial Z'_i(z(Z,t),t)/\partial Z_i=0$, 
as can be shown by differentiating both parts of (11) and 
using Cramer's rule.   Therefore the addition 
contributes entirely to the reversible drift,
reducing to the   Poisson bracket $[G,\Pi]$ where
\begin{equation}
\partial G(Z,t)/\partial X=-P', \quad\partial
G(Z,t)/\partial P=X',
\end{equation}  %   (12)
$(X,P)$ denote the conjugated $n$-component  
coordinates and
moments of $Z$, and $(X',P')=Z'$.   That is,
\[\partial \rho(z,t)/\partial t -[H,\rho]=
\partial \Pi(Z,t)
/\partial t-[\mathrm{H},\Pi] \]
with both sides understood as functions of 
either $z$, $t$ or $Z$, $t$ and where $\mathrm{H}$ is 
the transformed Hamiltonian,
\begin{equation}
\mathrm{H}=H+G.
\end{equation}  %   (13)
So, nothing is added  to the 
$\partial S_i/\partial z_i$ part 
of (2) under the transformation and this part preserves 
invariance. This and that $\rho$ is canonically 
invariant completes the proof of the theorem if $S$ is 
purely irreversible, i.e. for the Hamiltonian $H$ 
implied dressed.  
The proof holds also for the part  
$\partial S_i/\partial z_i$ decomposable into a pure 
irreversible term plus Poisson bracket 
terms. The latter are canonically invariant, so is 
therefore the irreversible term and hence its operator. 

It follows from the theorem that the time-symmetry 
decomposition once 
adopted in (2) must hold further for an arbitrary 
time dependence of the parameters of $H$ and $S$. 
The irreversible operator, being invariant 
under arbitrary canonical transformations and  
initially adopted related to the system's
motion given by the invariant Poisson bracket 
$[z, H]$, must preserve its dependence on $H$, while 
the reversible operator must change as $H$ transforms 
into $\mathrm{H}$.  This means the energy duality. 
The difference $G$ between the two Hamiltonians  
depends on the  $Z'$ given by (12) and can much 
exceed $H$ in effect. Respectively the effects of 
energy duality can be strong. 
 
In general the  decomposition of $S$ is in question.  
Say, for $S$ representing a retarded functional 
of $\rho$, i.e., acting not only in argument $z$ but 
also in $t$, the notion of irreversible operator  
is vague and the theorem loses its force. As for the 
energy duality principle, it is to be viewed 
applicable in compliance with theorem III  to an 
initially adopted $H$ versus the  
operator of whole $S$.  

\subsection*{Averaging versus canonical transformations}
The emergence of irreversible behaviors is often referred  
to the averaging of a conservative many-body system 
given by a microscopic Hamiltonian and random initial 
conditions.  
This cue misleads   in the question of both statistical and 
dynamical averaging.  
Averaging the Liouville's equation of the system 
over the statistics of initial conditions specifies 
the initial distribution function while its probabilistic 
evolution is governed by the same equation. Also,
so long as its solution is unique, integrating this equation  
over the irrelevant variables reduces to a 
canonical transformation splitting the microscopic 
Hamiltonian into an averaged part independent of the 
remaining 
part, which leaves no place for the irreversible 
probability currents. 

While splitting of a Hamiltonian system into an averaged 
subsystem independent of the rest corresponds, ideally,    
to some canonical transformation, thinking the same 
way of systems where irreversible flows matter is incorrect.
No questions arise for the averaging over the interactions 
and influences treated as small, disregarding their 
nonlinear effect; but the averaging causing appreciable 
irreversible flows is different, and it never reduces 
to  canonical transformations, for by virtue of theorem 
IV such flows would not emerge or change. 
This is so whether the system is conservative or 
not. 
As for the systems under time dependent influences treated 
as a random process, its exact statistical 
averaging causes irreversible flows as set already on the 
probability evolution of the random process.  
Some way or other,   
the irreversible flows emerge or change only due to 
truncations of interactions. Just as the flows, the energy 
duality is a relative category. 
Regularly recurring processes lend weight to them, and 
 so is the scientific cognition. 

Practically,  a theory established for some conditions 
is extended further to interactions with more fields to 
model transitions to underspecified states and 
instabilities. 
The trend of irreversible flows then may change, act as 
inverse truncation if left intact,  
and, being uncertain, it is usually assumed in 
accordance with the paradigm of entrainment ideal. 
For example, the impact of fields associated with 
Feshbach resonance [5]  on  many body systems is 
commonly treated presently as  the initial entrainment 
ideal $\rho_\mathrm{st}=\rho_\mathrm{st}(H)$ turning 
into the final ideal
$\rho_\mathrm{st}=\rho_\mathrm{st}(\mathrm{H})$ 
where Hamiltonian $\mathrm{H}$ includes  the field. 
Departure from this taken-for-granted trend  
in transitions means admitting  a steady non-entrained 
state, a relaxation sort of persistent currents setting in. 

An important field where at stake is whether 
to stick to the generalized thermodynamic potential 
or follow  the principles of energy duality is 
the impact of high frequency fields on  the system's 
probability  evolution. So we can view also the foregoing 
persistent currents  in the moving picture where they 
look frozen and where the arising hf field is to be 
treated self-consistently.  
Let us dwell here on typical models of 
$H$ containing hf terms treated as given and  $S$ of
form (3) with the constraints on $d$ and $f$ as stipulated
in the paragraph with Eq.~(9)  for such $H=H(z,t)$. 
The  hf part of $H$ is assumed
of finite amplitude and frequencies high enough  to
invalidate the approximation of distribution $\rho(z,t)$
by quasistationary Boltzmann distribution (6). 
A characteristic measure often in use then, and we 
question its extensive use, is 
the Boltzmann $\rho(z,t) = \Pi_{_\mathrm{B}}[Z(z,t)]$,
\begin{equation}
\Pi_{_\mathrm{B}}(Z) = e^{-\beta \mathrm{H}(Z)}\left[\int
e^{-\beta \mathrm{H} (Z)} d^{2n}Z \right]^{-1},
\end{equation}  % (14)
in terms of canonical $Z$ where  $\mathrm{H}(Z)$ is the
quasienergy, the $t$-independent function $\mathrm{H}$
given by (13), or its surrogate given by the 
effective potential i.e. the Hamiltonian of smoothed
dynamics $Z(t)$ obtained on hf averaging the equations
$dz/dt=[z,H(z,t)]$ for $z=z(t)$.

The conversion to terms of quasienergy or effective
Hamiltonian is a highway in physics, it clears up the
energy considerations of purely dynamic systems since
the $t$-dependence of $H$ makes their energy  a vague
notion. However, the measure (14), being the  exact
steady distribution for zero hf field,  violates  the
canonical invariance theorem  and leaves no place
for the energy duality  for any finite hf fields,
which suggests  the inadequacy of (14) in many 
conditions.

Obviously the approximation (14) is not valid on the
stages before thermalization unless the state is prepared
so initially. But even prepared so, the dynamic system
$Z(t)$ given by the Hamiltonian  $\mathrm{H}(Z)$ is not
entrained. As it vibrates with respect to the environment
persistently and rapidly, in time with the hf field,
there  arise  mean vortex forces having  a cumulative
bearing  on the system's behaviors which is  generally
substantial,  as elucidated below.

\subsection*{The vortex impact of time dependent fields}
The division of generalized forces and
impacts into types implies reasoning and transforms
in terms of specific variables. Canonical  ones
give division with respect to irreversibility, and
the terms where the system Hamiltonian is independent
of time give a subdivision -- of reversible forces
into potential versus gyroscopic and, as shown below,
of their irreversible counterpart into vortex  versus
non-vortex forces.

Transforming Eq.~(2) with  constraints  (3), (9)
where $H=H(z,t)$ includes given ($t$ dependent) fields  
to canonical variables
$Z$ where $\mathrm{H}$ (13) is $t$-independent, we obtain
\begin{equation}
\frac{\partial}{ \partial t}\Pi=[\mathrm{H},\Pi]
+\frac{\partial}{
\partial Z_i}\left(\beta D_{ik}\frac{\partial
(\mathrm{H}-G)}{\partial Z_k}+
D_{ik}\frac{\partial}{\partial Z_k}\right)\Pi
\end{equation}  %   (15)
where the elements
\begin{equation}
D_{ij}=d_{kn}\frac{\partial Z_k(z,t)}{
\partial z_i}\frac{\partial
Z_n(z,t)}{\partial z_j}
\end{equation}  %   (16)
comprise the  matrix $D$ of diffusion. $D$ is positive
semi-definite like matrix $d$ by virtue of (11). $\Pi$,
$G$ and $D$ are understood in (15) as functions of $Z$
and $t$, with $z$ expressed via $Z$ and $t$. Obviously
both $z\rightarrow Z$ and $Z\rightarrow z$ are one-to-one 
maps by virtue of (11).

Since  $\mathrm{H}$ of (15) is independent of $t$, the
components
\begin{equation}
F^\mathrm{ei}_i=
-\beta D_{ik}\frac{\partial \mathrm{H}}{\partial Z_k}
\end{equation}  %   (17)
(where  $i,k=1,\ldots 2n$)  comprise the irreversible drift
force $F^\mathrm{ei}$ corresponding to the entrainment 
ideal,
and the irreversible drift force $F^\mathrm{ne}$ of 
components
\begin{equation}
F^\mathrm{ne}_i=\beta D_{ik}\frac{\partial G}{\partial Z_k}
\end{equation}  %   (18)
is due to the non-entrainment caused by the given field.

Let us juxtapose the work done by these two forces and
diffusion on the system. Multiplying both sides of (15) by
$\mathrm{H}$ and integrating over  space $Z$ results in
\begin{equation}
\frac{d\langle \mathrm{H}\rangle}{ dt}=-\beta\langle
D_{ik}\bar{V}_i\bar{V}_k \rangle+\beta\langle
D_{ik}\bar{V}_i\bar{Z'}_k \rangle +\langle \frac{\partial
D_{ik}\bar{V}_i }{\partial Z_k}\rangle
\end{equation}  %   (19)
where $\bar{V} = \partial \mathrm{H}/\partial Z$ is the
vector orthogonal to $[Z,\mathrm{H}]$ and its norm
$|\bar{V}|= |[Z,\mathrm{H}]|$, and $\partial G/\partial Z$
of (12) is rewritten as $\bar{Z'}$ since it is orthogonal
to $Z'$ and $|\bar{Z'}|=|Z'|$.

The quadratic form $ D_{ik}\bar{V}_i\bar{V}_k\geq 0$,
hence the power of force $F^\mathrm{ei}$ (17) is always
dissipative, tends to decrease $\langle\mathrm{H}\rangle$.
The  mean power of diffusion forces, the third sum in the 
rhs of (19), is always positive, anti-dissipative
near steady states in the limit of weak given fields, 
for only it then remains to balance
dissipation due to $F^\mathrm{ei}$.

In contrast, the  power of force $F^\mathrm{ne}$ can be 
positive as well as negative. It is positive  when the 
vectors
$\bar{V}$ and $\bar{Z}'$ are parallel and negative
when antiparallel. So the sign of  work of
$F^\mathrm{ne}$ over a closed path in phase space $Z$
depends on direction of path-tracing, which is
intrinsic of a mean vortex force. As evident from (19),
the mean power of vortex force  prevails over that of
dissipative $F^\mathrm{ei}$ as the speed $\bar{V}$ of 
system decreases compared to  
the speed $\bar{Z}'$ of non-entrainment.

Not only the steady distribution can then  strongly
differ  from Boltzmann (14), but also the stability
threshold of the system can appreciably retreat toward
as inside as outside the  domain given by (14), i.e.,
where the $\mathrm{H}(Z)$ is bounded below. Also
the rates of relaxation can be much higher or lower
than given by eigenvalues of $\beta D$.

Particularly this concerns the nonlinear impact of  
high frequency fields, even relatively weak, 
as is the case of long lasting rotation/vibration
systems near resonances, including parametric and
combinational. This was brought up in various fields
by the present author, e.g.,[6-12]. 
Note that  we
referred  to as vortex there  only the vortex force
field in the coordinate space of system, rather
than in all phase space, as in the present work. 

\subsection*{Multi-bath extension}
The energy duality formulated above is with respect 
to the guide of entrainment limit. It does not 
cover  the realm of  environments not 
carried along between themselves on the average. 
However, the energy duality principle then  
displays 
through treating the realm as a superposition of 
environments ("baths") each compliant with detailed 
balance and moving with respect to each other.  
Applied to diffusional approximation, this is 
a number of baths each compliant with constraints (9) 
on $d$ and $f$ in the picture of canonical variables 
where the system is carried along it. 
On account of the canonical invariance
theorem we then come out with the irreversible drift
force $f$ related to the diffusion matrix $d$ in Eq.~(3) 
so
\begin{equation}
f=-\sum\beta^\mathrm{r}d^\mathrm{r}_{ik}\frac{\partial
(H+G^\mathrm{r})}{\partial z_k},
\quad d=\sum d^\mathrm{r}.
\end{equation}  %   (20)
Here $\sum$ means summing over the superscript $r$
labeling the quantities related to each of the baths,
$\beta^\mathrm{r}$ is the noise-energy scale of
bath $r$, $d^\mathrm{r}$ its contribution  to the
diffusion rates of the system, and $
H+G^\mathrm{r}$
the Hamiltonian transformed to the picture of
canonical variables $z^\mathrm{r}$ at rest with
bath $r$. With $G^\mathrm{r}$ given in this picture
as function of $z^\mathrm{r}$ and $t$, we have
\begin{equation}
\frac{\partial G^\mathrm{r}(z^\mathrm{r}(z,t),t)
}{\partial
z_k}=\frac{\partial z_i^\mathrm{r}(z,t)}{\partial z_k}
\frac{\partial
\bar{z}_i^\mathrm{r}(z,t)}{\partial t}
\end{equation}  %   (21)
where  $ \bar{z}^\mathrm{r} 
=(-p^\mathrm{r},x^\mathrm{r})$,
$x^\mathrm{r}$ are the coordinates and $p^\mathrm{r}$ 
the 
 moments of $z^\mathrm{r}=(x^\mathrm{r},p^\mathrm{r})$.

It follows from (20) that the components
\begin{equation}
f^\mathrm{ei}_i=-\sum\beta^\mathrm{r}
d^\mathrm{r}_{ik}\frac{\partial
H(z,t)}{\partial z_k}
\end{equation}  %   (22)
comprise the irreversible drift force
$f^\mathrm{ei}$ exerted on the system 
in neglect of its motion relative to the baths, 
while the  force $f^\mathrm{ne}$ of components
\begin{equation}
f^\mathrm{ne}_i=-\sum\beta^\mathrm{r}d^\mathrm{r}_{ik}
\frac{\partial G^\mathrm{r}}{\partial z_k}
\end{equation}  %   (23)
with $\partial G^\mathrm{r}/\partial z_k$ 
given by (21) is  due to the relative motion of 
the system. The  
 $f^\mathrm{ne}$ is vortex in space $z$ and   
 $f^\mathrm{ei}$ is non-vortex, 
similar to the  hf forces
(18) and (17) in space $Z$. We now arrive in 
space $Z$ with the energy balance equation
\begin{eqnarray}
\frac{d\langle \mathrm{H}\rangle}{dt}=
&-&\sum
\beta^\mathrm{r}\langle D^\mathrm{r}_{ik}
\bar{V}_i\bar{V}_k\rangle
\nonumber \\&+&\sum \beta^\mathrm{r}\langle
D^\mathrm{r}_{ik}
\bar{V}_i(\bar{Z}'_k-\bar{Z}'^{\mathrm{r}}_k)
\rangle +\langle \frac{\partial D_{ik}\bar{V}_i }
{\partial Z_k}\rangle \hspace{10pt}
\end{eqnarray}  %   (24)
where $D^\mathrm{r}_{ik}$ is of form (16) with $d_{kn}$
replaced by $d^\mathrm{r}_{kn}$ and
 $\bar{Z}'^{\mathrm{r}}$ stands for 
$\partial G^\mathrm{r}/\partial Z  
=J\partial G^\mathrm{r}/\partial z$ with $J$ 
the inverse of matrix $\partial z(Z,t)/\partial Z$.

As with (19), the first sum on the right of (24) is always
dissipative, tends to decrease the mean energy $\langle
\mathrm{H}\rangle$ of the system and the second sum
represents the power of the mean vortex force whose work 
over
a closed path in phase space $Z$ depends on direction of
path-tracing. 

Remarkably, the relation (20) between the irreversible 
drift force and diffusion for  
$\partial G^\mathrm{r}/ \partial z= 0$ for all 
baths looks like  the constraint (9) with 
$\beta=\sum \beta^\mathrm{r}d^{-1}d^\mathrm{r}$. But, 
this $\beta$ is a matrix function of $z$ rather than a 
constant 
scalar.  
It is easy to show that the relation (9) with such 
$\beta$ represents a general constraint for the systems 
in the entrainment limit of diffusional approximation. 

It follows that    
the equipartition theorem in such conditions of 
entrainment 
generally does not take place and that the steady 
distribution $\rho_\mathrm{st}$  can be reduced to a 
superposition of a number ($\leq 2n$) of Boltzmann 
distributions. The latter 
form of solution breaks beyond the entrainment limit 
as well, which can be used in determining the limit.  

To illustrate features  of the energy duality and associated 
vortex physics arising in non-entrainment conditions, a simple 
example relevant to applications is considered below. 

\subsection*{The vortex
 confinement of particles in fields of rotational symmetry}
Here we apply our approach to long time confinement of 
particles with
fields that are constant in some rotating frame. 
For simplicity the particle trap formed in axial 
direction by
the field is assumed harmonic and independent of transversal
particle motion. The transversal force field is allowed  of
arbitrary azimuthal asymmetry. So are Penning and many other
traps for charged particles and neutral atoms, including 
a general quadrupole type of traps suggested by the author 
in 
90-ties, see references in [12] devoted to  elemental vortex 
statistical mechanics of relevance. 
While the theory of generalized thermodynamic potential
gained  ground in such traps [13], it seems to be severely 
restricted in view of non-entrained background gas, noisy 
field sources and  retarded reaction of trap field system 
(due to its finite conductance)  to the particle motion. 

Let us rely on [12], bringing in a wider insight and 
dwelling on the probability of particle motion governed  
 by Eqs. (2), (3) with constraints of  form (4) for a 
 number of baths each rotating
 about the symmetry axis of trap at some frequency
 $\Omega^\mathrm{r}$ and of constant, isotropic
 diffusion rates $d^\mathrm{r}$. In the rotating-frame 
coordinates where the trap field is 
 independent of  $t$, the kinetic equation (15) for 
 the probability distribution of particle states 
 reduces to
\begin{equation}
\frac{\partial}{\partial t}\Pi(X,P,t)
=[\mathrm{H},\Pi]+\frac{\partial}{\partial P_i}
\left( \beta D V_i -F_i^\mathrm{vort}
+D\frac{\partial}{\partial P_i}\right)\Pi
\end{equation}   %   (25)
where 
\[D=\sum d^\mathrm{r} \quad \text{and}\quad \beta
=\frac{1}{D}\sum \beta^\mathrm{r}d^\mathrm{r}.
\]
$V=\partial \mathrm{H}/\partial P$,   $-\beta DV$ is 
the friction force $F^\mathrm{ei}$ (17),   
 $F^\mathrm{ne}$ (18) takes the form of net vortex 
 force $F^\mathrm{vort}$ in space $X$. For a particle 
 of  Euclidean coordinates  $X=(X_1,X_2,X_3)$ 
 \begin{equation}
F^\mathrm{vort}=(\kappa X_2,-\kappa X_1,0),
\quad \kappa=\sum
(\Omega-\Omega^\mathrm{r})\beta^\mathrm{r}
d^\mathrm{r},
\end{equation}  %   (26)
$\Omega$ is the rotation frequency and $X_3$ along 
the rotation axis of trap. 
  In the trap field modeled 
as given, the Hamiltonian of  particle 
motion  is of form 
 \begin{eqnarray}
\mathrm{H}&=&\frac{1}{2}[P^2 +k_3X_3^2+
(k_0-2q)X_1^2+(k_0 +2q)X_2^2]\nonumber \\
&+&(g-\Omega )M_\Theta + \mathrm{H}^\mathrm{nl}
\end{eqnarray}  %   (27)
where $P=(P_1,P_2,P_3)$ are canonical moments
conjugated to $X$. $X,P$ are related
to the canonical $x,p$ of rest frame so
$X=T(t)x$, $\quad P=T(t)p$, 
\[\quad T(t)=\left(%
\begin{array}{ccc}
  \cos\Omega t & \sin\Omega t & 0\\
  -\sin\Omega t & \cos\Omega t & 0\\
  0 & 0 & 1\\
\end{array}%
\right), \]
$G=\mathrm{H}-H=-\Omega M_\Theta$ where 
 $H$ is  the rest-frame Hamiltonian and     
$M_\Theta$  the canonical angular momentum
\begin{equation}
M_\Theta=x_1p_2-x_2p_1=X_1P_2-X_2P_1,
\end{equation}  %   (28)
$k_0=g^2-k_3/2$, $k_3>0$, $\sqrt{k_3}$ is the 
eigenfrequency  of axial oscillator mode, $q$  
the gradient of  rotating transversal quadrupole 
potential and $\mathrm{H}^\mathrm{nl}$ the  
potential of transversal multipoles higher than 
quadrupole. $\mathrm{H}^\mathrm{nl}=0$ for the 
geometries of Penning and rotating quadrupole traps. 

The confinement in line with the theory of  generalized
thermodynamic potential corresponds to the ideal of 
$\kappa=0$ and would mean that the particle should relax 
to the entrained state of rotating-frame Boltzmann 
distribution (14) in the 
parameter  domain where  $\mathrm{H}(Z)$ (27) is bounded 
below, and  an instability should arise 
 as the boundedness breaks down; 
for  $\mathrm{H}^\mathrm{nl}=0$, it breaks down where
\begin{equation}
k_0<(\Omega-g)^2.
\end{equation}  %   (29)
The critical point $k_0=(\Omega-g)^2$ is the apex of 
parametric resonance of transversal motion caused by the 
rotating quadrupole. 
However, the trends of long time 
confinement in considered traps strongly diverge from 
these predictions. For example, the traps of $g=0$ would 
be impossible in principle, for any $q$, if condition 
(29) would indeed imply instability, but it may 
not at all in view 
of the non-entrainment  given by $\kappa\neq 0$. 

An important point to account for $\kappa\neq 0$ 
is that the work of vortex force over a 
closed path enclosing in a transversal plane 
(across axis $X_3$) an area 
$s$ is equal to 
$\oint F^\mathrm{vort}dX =\pm 2\kappa s$, 
velocity independent, and the sign depends on 
the direction 
of motion, while the work $\oint \beta DV dX$ decreases 
with $V$ and is of one sign. For a circular transversal 
motion of a frequency $\omega$ 
\begin{equation}
\frac{\oint F^\mathrm{vort} dX_\bot}{\oint -\beta 
DV dX_\bot}=
\frac{\Omega^*}{\hspace{-3pt}\omega},
\quad \Omega^*=\Omega-
\frac{\sum \Omega^\mathrm{r}\beta^\mathrm{r}
d^\mathrm{r}}{ \sum
\beta^\mathrm{r}d^\mathrm{r}}
\end{equation}  %   (30)
where $X_\bot$ denotes a vector $(X_1,X_2)$. 
A sizable proportion of $\Omega^*$ is due to the 
environments
roughly at rest with the trap electrodes, of 
$\Omega^\mathrm{r}\ll\Omega$, 
so $\Omega^*$ and $\Omega$ are within 
 one order. The soft mode of transversal 
dynamics given by $\mathrm{H}$ is slow, of  
frequency scale 
$\nu\ll \Omega$. So $\omega\ll\Omega^*$ for such modes.
The  particle stability, relaxation and steady state is 
critical to the soft mode, hence the dominance of vortex 
force impact on all that.

Importantly the dominance holds on passing the critical 
point into a certain parameter domain  of $\mathrm{H}(Z)$ 
unbounded from 
below, where the soft mode still represents a transversal 
orbiting, rather than runaway motion. Such a domain is 
characteristic of dynamic systems with gyroscopic terms, 
as 
is the case of particle traps with rotating fields. For  
$\mathrm{H}^\mathrm{nl}=0$, the Hamiltonian of 
transversal dynamics presented in normal mode 
presentation in the domain takes the form [12]
\begin{equation}
\mathrm{H}_\perp=\frac{1}{2}\nu_+(P_+^2+X_+^2)
\pm\frac{1}{2}\nu_-(P_-^2+X_-^2)
\end{equation}      %   (31)
where '$\pm$' is  '$+$' for $k_0>(\Omega-g)^2$ and '$-$'
for $k_0<(\Omega-g)^2$, $X_+,P_+$ are canonical 
variables
of normal '$+$' mode  and $X_-,P_-$ of '$-$' mode, 
$\nu_\pm$ are positive roots of
\begin{equation}
\nu_\pm^2=k_0+(\Omega-g)^2
\pm 2\sqrt{(\Omega-g)^2k_0+q^2},
\end{equation}  %   (32)
and the domain of no runaway is where
\begin{equation}
\nu_-^2>0 \quad\mathrm{and}\quad k_3>0.
\end{equation}  %   (33)
For the apex falling in domain (33), the '$-$' mode  is 
soft near the apex and its energy is negative below it.

Along with the dominance, another important feature of 
vortex force displays vigorously and not alike 
the friction force $-\beta DV$.  The latter pumps energy 
out of any motion, so its
pumping out of positive-energy modes causes their
damping, and pumping out of negative-energy  
modes  causes oscillation build-up, a negative friction.  
So,  on accounting only for the fiction force, the 
condition 
(29) would imply instability of increment $\sim \beta D$. 

In contrast, the stabilizing/destabilizing impact of the
vortex force  does not change sign on passing the critical
point but changes it on passing  the point of gyroscopy 
compensation $g=\Omega$ by field rotation, unless  both 
points coincide.  Indeed, the energy pumped out/in  
the soft
mode by the vortex force depends on  the ellipticity  
of transversal-mode orbiting  and whether in the 
rotating frame the orbiting is  in  the direction of 
field rotation or opposite to it. For  $g\neq\Omega$, 
the sense of  orbiting changes twice, on passing both 
points, but on passing the critical point there changes 
the sign of energy transfer
by the vortex force, hence, its 
stabilizing/destabilizing 
impact does not change sign on transition into 
the domain 
of $\mathrm{H}(Z)$ not bound  below for the same reason 
why there changes the sign of friction force impact.

For the models considered in this section, the vortex 
impact vanishes for the transversal modes of linear 
polarization, which is when $g=\Omega$, and is maximal 
for circular orbiting. For $\mathrm{H}^\mathrm{nl}= 0$, 
calculating the time averages
of energy flows due to the vortex and friction forces in 
the orbiting motion corresponding to the '$+$' and 
'$-$' modes  with exact account of their polarization 
forms, 
we obtain for their ratio, respectively
\begin{equation}
\left(\frac{\oint F^\mathrm{vort} dX_\bot}{\oint
-\beta DV dX_\bot}
\right)_\pm=\frac{\Omega^*}{\hspace{-3pt}\Omega}
\frac{(g-\Omega)\Omega}{\nu_\pm^2-k_0}.
\end{equation}  %   (34)
The ratio $\Omega^*/ \Omega>0$, so the sign of 
ratio next to $\Omega^*/ \Omega$ in (34)  determines 
the sign of stabilization/destabilization impact of 
the vortex force. There the dominator $\nu_\pm^2-k_0$ 
changes sign only 
on passing the  apex for the '$-$' mode, as evident 
from (32). Thus the  expressions (34) and (33) show 
all variety  of  confinement trends in question.

For $g=0$, for example, $k_0<0$ and the domain (33) is  
completely on the side of apex where (29) holds and the 
soft mode is of negative energy. However, the vortex 
force exerts there stabilizing action on the soft mode, 
and for $\Omega^*=\Omega$ it prevails over destabilizing 
effect of  force $-\beta DV$ in \emph{all} domain (33). 
As for the  '$+$' mode, the vortex impact on it is 
destabilizing but weaker than frictional which is  
stabilizing. So,  a stable 3D confinement takes place in 
all domain  (33)!

At the point of gyro-compensation $g=\Omega$, the 
vortex factor (34) vanishes and the trends of 
confinement in the trap fields of 
$\mathrm{H}^\mathrm{nl}= 0$ comply with the theory 
of generalized thermodynamic potential. However, this 
specific point is aside from conditions most favoring 
particle cooling and confinement in traps, as well as 
heating and selection of particles, and there the 
vortex force rules the trends.  Depending on the ratio 
$g/\Omega$ and its sign, the vortex impact appears to 
be either stabilizing or destabilizing, resulting in  
amazing trends of stability, relaxation rates and 
steady states.

Highly important in all that is the scale 
$\Omega^*/ \Omega$, whether it is small enough to stick 
to the entrainment-ideal scenario or large, comparable 
with 1, to have an essentially vortex picture of 
confinement. Characteristically the scale 
$\Omega^*/ \Omega$ and hence the factor of prevalence 
of the vortex impact over frictional is the same  for 
any positive values of baths' parameters $\beta$ 
and $D$, 
including  the limit of $\beta D\rightarrow 0+$.

For $\mathrm{H}^\mathrm{nl}= 0$, the kinetic equation 
(25) admits exact analytical solution  $\Pi(Z,t)$. It 
represents, for the system evolution  from a given 
initial state in a point of phase space $Z$, a 
multivariate Gaussian distribution of system states 
for any $t>0$. So  all probability characteristics of 
transient and steady states can be traced exactly for 
the entire range from $\Omega^*/\Omega=0$ to 1.

For $\mathrm{H}^\mathrm{nl}\neq 0$ the  transversal 
Hamiltonian as nonlinear is generally inseparable into 
normal modes, but in  any close parameter vicinity of 
inseparable Hamiltonian there exist separable  ones. 
So for fairly long times the motion in the domain of 
no runaways can be treated as separable and of soft 
mode on approaching apexes of  parametric or combination 
resonances. Therefore the vortex 
trends resembling those shown above are to be expected.

\paragraph*{Additional remarks.}
For more complex objects than a 3D-particle  in a 
field of rotational symmetry, representing nonlinearly 
coupled systems having many degrees of freedom of high 
and low frequencies,
the vortex impact appears to be no less significant, 
as is, e.g.,  in conditions of  hf subsystems excited 
near  main or parametric or combinational resonances. 
Its general features in the lowest order of 
nonlinearities 
were elucidated in [6-12] where  various methods of cooling 
and control of interacting particles, waves, domain walls 
etc were suggested on this basis. 

Imposing on a  system in  conditions of entrainment  
a regularly varying field modeled via Hamiltonian 
interactions always translates into a vortex impact 
exerted  on the system in the picture where the field 
looks frozen.  This refers to any control of irreversible 
effects this way.
So works the principle of energy duality.  Now, with this 
principle, there arises a consistent and adequate energy 
approach to the above-cited and many other phenomena in 
open systems.

\subsection*{Conclusion}
We have shown that  the notion of energy function developed 
in analytical mechanics and thermodynamics acquires rigor 
and significance not only in the limit of detailed balance 
but also in the statistical approach to the vast world where
it rules relaxation to non-entrained states, and that this 
comes with breaking the energy function into two having 
different transformation rules. 

The two functions merge 
in the entrainment ideal and can strongly differ beyond it. 
As demonstrated, both are important for the energy 
approach to the long time behavior and stability of systems. 
Otherwise, huge inconsistencies arise, as is the case of 
particle confinement and other systems mentioned in 
this work.  

Behind the energy duality there stands the persistent 
irreversible drift related   to the difference 
between the two energy functions of system and 
representing a vortex force field  in its phase space. 
It is  essential that the vortex force effect  differ 
radically from that of familiar friction of systems
relaxing to entrained states. We see its distinctive 
features 
as in the  introduced classification of generalized 
vortex forces and energy flows as in the physics the 
approach gives to various phenomena. 

The principle of energy duality does not rely on quantum 
mechanics and one cannot but infer from  the correspondence 
principle that a consistent concept of energy quanta is 
to be either  limited by the entrainment ideal or the 
energy 
quanta break into two respective sorts beyond the ideal.  
This  disputes common reasonings of 
non-entrained states in terms of energy quanta.  

While the generalized thermodynamic potential theory
fits only the entrainment limit, it may look 
all-sufficient 
beyond it with fitting via extending the set of variables 
and parameters.  However, the theory then loses its 
predictive strength and causes misleading associations like 
the ones revealed in this work.  The energy duality is free 
of these drawbacks and presents a universal energy concept 
incorporating the vortex physics.

\end{document}